\documentclass[a4paper,11pt]{article}

\usepackage{jheppub} 

\title{\boldmath Localization on round sphere revisited}

\author{Akinori Tanaka}
\preprint{OU-HET 794}

\affiliation{Department of Physics, Graduate School of Science, 
\\
Osaka University, Toyonaka, Osaka 560-0043, Japan}

\abstract{
We consider supersymmetric gauge theories on \textit{round} 3-sphere with a certain \textit{background gauge field}.
The Lagrangians break the usual symmetry because the background gauge field which we have turned on violates the isometry.
In order to maintain the supersymmetry, we choose unfamiliar charged Killing spinors as $\mathcal{N}=2$ SUSY parameters.
We perform localization calculous within this setup and find the double sine function as we expected.
We comment on more direct relationship between theories on \textit{round sphere} and \textit{squashed sphere}
via Weyl transformation.
}

\newcommand{\ba}{\begin{alignat}{3}}
\newcommand{\e}{\epsilon}

\newcommand{\dl}{\delta}
\newcommand{\g}{\gamma}

\newcommand{\tr}{\text{Tr}}

\newcommand{\pa}{\partial}

\newcommand{\om}{\omega}

\newcommand{\si}{\sigma}

\newcommand{\D}{\mathcal{D}}

\newcommand{\lam}{\lambda}
\newcommand{\olam}{\overline{\lambda}}
\newcommand{\oep}{\overline{\epsilon}}
\newcommand{\ttheta}{\tilde{\theta}}

\newcommand{\ophi}{\overline{\phi}}
\newcommand{\opsi}{\overline{\psi}}
\newcommand{\oF}{\overline{F}}

\newcommand{\Ai}{\mathcal{A}}

\begin{document} 
\maketitle
\flushbottom

\section{Introduction}\label{intro}
Exact calculations based on supersymmetric localization principle with various manifolds \cite{Pestun:2007rz, Kapustin:2009kz, Hama:2010av, Jafferis:2010un, Benini:2012ui, Doroud:2012xw, Hama:2011ea, Imamura:2011wg, Gang:2009wy, Kallen:2011ny, Ohta:2012ev, Alday:2013lba, Nian:2013qwa} are well studied in these days.
In particular, calculation on 3-dimensional sphere ($S^3$) is important for understanding of M2-brane dynamics.
In order to construct supersymmetry, it is necessary to fix a metric on $S^3$ because we should have a spin structure.
As the simplest way for it, authors in \cite{Kapustin:2009kz} have considered round $S^3$ and have got  hyperbolic functions as integrand for resulting matrix model.
After that, these results have been applied to ABJM theory \cite{Aharony:2008ug, Drukker:2010nc, Fuji:2011km, Marino:2011eh, Klemm:2012ii}, and other less supersymmetric theories \cite{Imamura:2008nn, Jafferis:2008qz, Herzog:2010hf, Santamaria:2010dm, Martelli:2011qj, Marino:2012az, Farquet:2013cwa}.
In \cite{Hama:2011ea, Imamura:2011wg}, these evaluations were generalized to those of on squashed $S^3$.
They have got so-called duoble sine functions after the quantum calculations.
At first sight, it might looks the squashed metric plays a crucial role for getting such one-parameter deformed functions.
However, this duoble sine functions are re-discovered within a broader class of metric on $S^3$ in \cite{Alday:2013lba}.
This fact suggests the possibility for getting same results on \textit{squashed} $S^3$ even on \textit{round} $S^3$ because the most important factor for deforming the results is a contact structure that comes from a background field.

Standing on this point of view, 
we construct SUSY gauge theories on round $S^3$ with a background field by using unusual Killing spinors which satisfy the following equations
\begin{align}
& \D_\mu \e = \g_\mu \Big( \frac{i}{2sf^2} \e - \frac{1}{2} \frac{f'}{f} \oep \Big) , \ \ \ 
\D_\mu \oep = \g_\mu \Big( \frac{i}{2sf^2} \oep - \frac{1}{2} \frac{f'}{f} \e \Big), \label{intro1}
\end{align}
and find the agreement with one-loop determinant \textit{on squashed sphere} \cite{Hama:2011ea, Martelli:2011fu} as we expected.

In addition, we examine 1/2 BPS condition for the supersymmetric Wilson loop
and find that it exists if and only if the parameter $s$ is a rational number.
In this case, the contour forms so-called torus knot that wraps a torus embedded in round sphere.
The expectation value for Wilson loop weighted by supersymmetric Chern-Simons action turns to be knot invariant expressed by matrix integral as discussed in \cite{Tanaka:2012nr}.

We follow the conventions \cite{Hama:2011ea} for bilinear products of spinors and gamma matrices.
But there are two differences. 
First, we treat Killing spinors $\e, \oep$ as Grassmann even and depending only on coordinate $\theta$ when we calculate SUSY algebra and Lagrangians.
Second, we take a gauge condition for the background $U(1)$ gauge field as same form in \cite{Martelli:2011fu}.

\section{Killing spinors on round three-sphere} \label{sec2}
We use following metric in this paper.
\begin{align}
ds^2 = d \theta^2 + \cos^2 \theta d \phi^2 
+ \sin^2 \theta d \chi^2 , \label{eq:round}
\end{align}
where the ranges of the coordinates are $\theta \in [ 0 , \frac{\pi}{2} ] $,  $\phi ,\chi \in [0, 2\pi]$.
Here $\phi, \chi$ are coordinates for the embedded torus, and periodic with period $2 \pi$. 
In order to construct supersymmetry on this curved space, let us consider the spinors
\begin{align}
\e = 
\begin{pmatrix}
-(\cos \theta + i s \sin \theta)^{1/2} \\
(\cos \theta - i s \sin \theta)^{1/2}
\end{pmatrix},
\quad
\oep=
\begin{pmatrix}
(\cos \theta + i s \sin \theta)^{1/2} \\
(\cos \theta - i s \sin \theta)^{1/2}
\end{pmatrix}.
\label{eKillings}
\end{align}
These spinors satisfy the following less familiar Killing spinor equations
\begin{align}
& \D_\mu \e
= \g_\mu \Big( \frac{i}{2sf^2} \e - \frac{1}{2} \frac{f'}{f} \oep \Big), 
\quad
 \D_\mu \oep
= \g_\mu \Big( \frac{i}{2sf^2} \oep - \frac{1}{2} \frac{f'}{f} \e \Big)
,
 \label{Killings}
\end{align}
where $f$ is defined by $ f^2 (\theta) = \sin^2 \theta + \frac{1}{s^2} \cos ^2 \theta$.
Note that $\e$ and $\oep$ are mixed in each right hand side.
We use here the following vielbein
\begin{align}
e^1 = \cos \theta d \phi,
\quad
e^2 = \sin \theta d \chi,
\quad
e^3 = d \theta.
\end{align}
The covariant derivative $\D_{\mu}$ is constructed from the round sphere spin connection $\om_\mu ^{ab}$, a background $U(1)$ field $V$ and its charge $\hat{q}$ as follows
\begin{align}
\D_{\mu} = \pa_{\mu} + \frac{1}{4} \om_\mu ^{ab} \g^{ab} - i \hat{q}  V_{\mu}, \label{covariant}
\end{align}
where $V$ is 
\begin{align}
V = V_{\mu} dx^{\mu} = \frac{1}{2f} (d\phi - \frac{1}{s} d \chi). \label{bg}
\end{align}
For example, $\e$ has $\hat{q} = +1$ and $\oep$ has $\hat{q} = -1$, etc.
We summarize formulas with Killing spinors and charge assignments for fields in Appendix \ref{A}.

\paragraph{A relation with known geometry}
The metric \eqref{eq:round} is known to be transformed to so-called squashed sphere metric \eqref{sqmet}\footnote{The geometry is slightly different from the ellipsoid considered in \cite{Hama:2011ea}, however results are equivalent. Therefore we would like to call it squashed sphere. In \cite{Martelli:2011fu}, they call it \textit{hyperbolic ellipsoid}.}  up to Weyl rescale \cite{Martelli:2011fu} by the change of coordinates
\begin{align}
 s \cos \ttheta = \frac{\cos \theta}{f(\theta)} ,
\quad
\tilde{\phi}  =\phi ,
\quad
\tilde{\chi}  = \chi,
\label{coordtr}
\end{align}
via
\begin{align}
d\tilde{s}^2 = (\frac{1}{sf(\theta)})^2 ds^2
  . \label{Weyl}
\end{align}

\paragraph{Relations with known Killing spinors}
Our unusual Killing spinors in \eqref{eKillings} themselves do not enter into the known classification studied in \cite{Closset:2012ru}.
However we can relate them with the known Killing spinors as follows.

By using squashed sphere metric \eqref{sqmet},
the spinors 
\begin{align}
\tilde{\e}=
\begin{pmatrix}
- e^{+\frac{i\tilde{\theta}}{2}} \\
e^{-\frac{i\tilde{\theta}}{2}} 
\end{pmatrix}
,
\quad
\tilde{\oep}=
\begin{pmatrix}
e^{+\frac{i\tilde{\theta}}{2}} \\
e^{-\frac{i\tilde{\theta}}{2}} 
\end{pmatrix}
\end{align}
satisfy the following well known Killing spinor equations \cite{Martelli:2011fu}
\begin{align}
\tilde{\D}_\mu \tilde{\e} = \frac{i h(\tilde{\theta})}{2} \tilde{\g}_\mu \tilde{\e},
\quad
\tilde{\D}_\mu \tilde{\oep} = \frac{i h(\tilde{\theta})}{2} \tilde{\g}_\mu \tilde{\oep}.
\end{align}
Here, each objects are defined by using \eqref{sqmet}.
Our unusual Killing spinors can be expressed by
\begin{align}
\tilde{\e} = (sf)^{-\frac{1}{2}} \e,
\quad
\tilde{\oep} = (sf)^{-\frac{1}{2}} \oep.
\label{WeylKilling}
\end{align}
In this sense, our spinors \eqref{eKillings} fall into the classification in \cite{Closset:2012ru} after we perform the Weyl transformation \eqref{WeylKilling}.

\section{Another SUSY on round $S^3$} \label{sec;theo}
We show a construction for $\mathcal{N}=2$ supersymmetry with round sphere metric  \eqref{eq:round},
 Killing spinors \eqref{Killings} and  background $U(1)$ gauge field \eqref{bg}.
Because our Killing spinor $\e$($ \oep$) depends on the parameter $s$,
$\D_\mu \e$ ($\D_\mu \oep$) splits into $\e$ and $\oep$.
Then SUSY transformation contains more extra terms compared with familiar cases  \cite{Alday:2013lba, Martelli:2011fu, Kapustin:2009kz, Hama:2011ea, Hama:2010av, Jafferis:2010un, Imamura:2011wg}.
\paragraph{Vector multiplets.}
Supersymmetry for fields in vector multiplet is as follows:
\begin{align}
&\dl _{\e} A_{\mu} = - \frac{i}{2} \olam \g_{\mu} \e  , 
\ \ \ \
\dl _{\oep} A_{\mu} =  - \frac{i}{2} \oep \g_{\mu} \lam ,  \label{Aoep} \\
&\dl _{\e} \si = + \frac{1}{2} \olam  \e   , 
\ \ \ \
\dl _{\oep} \si =  + \frac{1}{2} \oep  \lam ,  \label{sioep} \\
&\dl _{\e} \lam =  \frac{1}{2}  \g^{\mu \nu} \e F_{\mu \nu} - D \e + i \g^{\mu} \e \D_{\mu} \si + \frac{2i}{3} \si \g^{\mu} \D_{\mu} \e  , 
\ \ \ \
\dl _{\oep} \lam =  0 , \label{lamoep} \\
&\dl _{\e} \olam =  0  , 
\ \ \ \
\dl _{\oep} \olam =  \frac{1}{2}  \g^{\mu \nu} \oep F_{\mu \nu} + D \oep - i \g^{\mu} \oep \D_{\mu} \si - \frac{2i}{3} \si \g^{\mu} \D_{\mu} \oep , \label{olamoep} \\
&\dl _{\e} D =  + \frac{i}{2}  \D_{\mu} \olam \g^{\mu} \e - \frac{i}{2} [\olam \e , \si ] + \frac{i}{6} \olam \g^{\mu} \D_{\mu} \e  , 
\dl _{\oep} D = - \frac{i}{2} \oep \g^{\mu} \D_{\mu} \lam + \frac{i}{2} [\oep \lam , \si ] - \frac{i}{6} \D_{\mu} \oep \g^{\mu} \lam. \label{Doep}
\end{align}
Note $\dl_{\e}, \dl_{\oep}$ are purely fermionic transformations so that we treat Killing spinors as Grasmann even.
Because of this statistic choice, some transformations look different compared with \cite{Hama:2011ea, Hama:2010av}.
By a direct calculation, one can verify $\dl_{\e}^2 =\dl_{\oep}^2 = 0$.
Also, $\{ \dl_{\e} , \dl_{\oep} \}$ turns to be a sum of other symmetries\footnote{Translation, gauge transformation, Lorentz and R rotation as the same form in a literature \cite{Hama:2010av}.}.
Here we would like to comment on our formulation.
As discussed in \cite{Hama:2011ea, Hama:2010av},
there is a nontrivial constraint on the spinors $\e,\oep$ in order to close the supersymmetry.
Our Killing spinors \eqref{Killings} satisfy the constraint.
\paragraph{Matter multiplets.}
We show matter multiplet supersymmetry transformation.
\begin{align}
& \dl_{\e}  \phi =0,
\ \ \ \
 \dl_{\oep}\phi = \oep \psi , \\
& \dl_{\e}  \ophi = \e \opsi, 
\ \ \ \ 
\dl_{\oep}\ophi = 0 , \\
& \dl_{\e}  \psi = i \g^\mu \e \D_\mu \phi + i \e \si \phi + \frac{2 \Delta i}{3} \g^\mu \D_\mu \e \phi ,
\ \ \ \
\dl_{\oep}\psi =  \oep F , \\
& \dl_{\e}  \opsi =  \oF \e ,
\ \ \ \
 \dl_{\oep}\opsi = i \g^\mu \oep \D_\mu \ophi + i \ophi \si \oep + \frac{2 \Delta i}{3} \ophi \g^\mu \D_\mu \oep , \\
& \dl_{\e}  F=  \e (i \g^\mu \D_\mu \psi - i \si \psi - i \lam \phi) + \frac{i}{3} (2 \Delta -1) \D_\mu \e \g^\mu \psi,
\ \ \ \
 \dl_{\oep}F = 0, \\
& \dl_{\e}  \oF=  0 ,
\ \ \ \
 \dl_{\oep} \oF = \oep (i \g^\mu \D_\mu \opsi - i \opsi \si  + i \ophi \olam) + \frac{i}{3} (2 \Delta -1) \D_\mu \oep \g^\mu \opsi .
\end{align}
Even our statistic choice of $\dl_{\e} , \dl_{\oep}$ is purely fermionic, there are no difference with transformations in \cite{Hama:2011ea, Hama:2010av} because Killing spinors are always putted on the left side in each spinor bilinear products.
One can verify that $\{ \dl_{\e} , \dl_{\oep} \}$ becomes a sum of 
other symmetries and $\dl_{\e}^2 =\dl_{\oep}^2 = 0$.
\paragraph{Chern-Simons Lagrangian.}
The usual Chern-Simons Lagrangian 
\begin{align}
\mathcal{L} _{\text{CS}}
=
\tr \Big[ \frac{1}{\sqrt{g}} \varepsilon^{\mu \nu \lam} (A_\mu \pa_\nu A_\lam - \frac{2i}{3} A_\mu A_\nu A_\lam) - \olam \lam + 2 D \si \Big]
\end{align}
is SUSY invariant even within our setup.
\paragraph{Yang-Mills Lagrangian.}
However, the Yang-Mills Lagrangian is slightly different from the usual one as follows:
\begin{align}
\mathcal{L} _{\text{YM}}
=
(2sf) \tr \Big( &
\frac{1}{4} F^{\mu \nu} F_{\mu \nu}
+ \frac{1}{2} \D^i \si \D_i \si
+ \frac{1}{2} (\D_\theta \si + \frac{f'}{f} \si ) ^2
+ \frac{1}{2} (D + \frac{1}{sf^2} \si )  ^2
 \notag \\ & 
 + \frac{i}{2}  \olam \g^i \D_i \lam
 + \frac{i}{2}  \olam \g^\theta (\D_\theta \lam + \frac{1}{2} \frac{f'}{f } \lam) 
 + \frac{i}{2} \olam [ \si,\lam ] 
 - \frac{1}{4}\frac{1}{sf^2}  \olam \lam
 \Big),
\end{align}
where $i$ runs for $\phi,\chi$.

\paragraph{Matter Lagrangian.}
The matter Lagrangian is in the same situation.
It is also different from usual form as follows:
\begin{align}
\mathcal{L} _{\text{mat}}
=
(2sf^{2\Delta -1}) \Big(&
 \D_\mu \ophi \D^\mu \phi 
 +  \ophi \si^2 \phi \color{black}
+ i  \ophi D \phi 
+\ophi \si^2 \phi 
+i \frac{2\Delta-1}{sf^2} \ophi \si \phi
                          \notag \\ & 
- \frac{\Delta(2\Delta-1)}{2(sf^2)^2}  \ophi \phi   - \frac{\Delta  (2\Delta -1)}{2} (\frac{f'}{f})^2  \ophi \phi  + \frac{\Delta}{4} R  \ophi \phi 
                          \notag \\ & 
- i \opsi \g^\mu \D_\mu \psi
-i \frac{f'}{f} (\Delta - \frac{1}{2} ) \opsi \g^3 \psi
 + i \opsi \si \psi 
- \frac{2 \Delta -1}{2sf^2} (\opsi \psi) 
                          \notag \\ & 
+i \opsi \lam \phi
- i \ophi \olam \psi
+ \oF F
\Big) \label{matlag}
\end{align}
The most important fact is that $\mathcal{L} _{\text{YM}}$ and $\mathcal{L} _{\text{mat}}$ can be expressed as supersymmetric-exact forms
\begin{align}
&\mathcal{L} _{\text{YM}} 
=
- \dl_{\oep} \dl_{\e} \tr \Big( \frac{1}{2} \olam \lam - 2 D \si \Big) , \label{suder1} \\
&\mathcal{L} _{\text{mat}}
=
- f^{2\Delta-2} \dl_{\oep} \dl_{\e}  \Big(  \opsi \psi - 2i \ophi \si \phi + \frac{2(\Delta-1)}{sf^2} \ophi \phi \Big), \label{suder2}
\end{align}
or, the following forms are easier to calculate
\begin{align}
&\mathcal{L} _{\text{YM}} 
=
- \dl_{\oep}  \tr \Big( \frac{1}{2} ( \dl_{\oep} \olam) \Big| _{\oep \to \e} \olam \Big) , \label{suder1b} \\
&\mathcal{L} _{\text{mat}}
=
- f^{2\Delta-2} \dl_{\oep}   \Big( \dl_{\oep} (\oF \phi) \Big|_{\oep \to \e}  \Big). \label{suder2b}
\end{align}
The subscript $\oep \to \e$ means replacing spinor in the supersymmetry transformations, for example
\begin{align}
\dl_{\oep} \phi \Big|_{\oep \to \e}
=
\oep \psi \Big|_{\oep \to \e}
=
\e \psi.
\end{align}
Note that $\oep = - \g_3 \e$.

\paragraph{Comments on Weyl transformations}
As explained, we can regard our system as Weyl rescaled squashed sphere,
and the above Lagrangians are exactly Weyl equivalent to the Lagrangians on the squashed sphere through redefining fields correctly as performed in \cite{Pestun:2007rz, Nagasaki:2011sh} in 4-dimensional case.
Therefore, it is natural that partition function and expectation values for BPS operators turn to be the same ones in \cite{Martelli:2011fu, Hama:2011ea}.

However, what we would like to emphasize in this paper is the capability for constructing supersymmetry and performing localization calculations which give us the same results on \textit{squashed sphere} within \textit{round sphere} metric.
This fact suggests that the metric is not so important for getting one-parameter deformed theories on $S^3$.

\section{Localization on round $S^3$} \label{sec;part}
From now on, we show how to perform the supersymmetric localization calculous on round $S^3$.
The field configurations 
\begin{align}
A_\mu = \phi = 0, \ \ \ \si = \frac{\si_0}{f}, \ \ \ D= - \frac{\si_0}{sf^3} , \label{locus}
\end{align}
called locus, vanish the bosonic part of the $\dl_{\oep}$-exact Lagrangians\footnote{We will give the definition for $\mathcal{L}_{\text{reg}}$ in \eqref{regm}.} 
$\mathcal{L} _{\text{YM}} , \mathcal{L}_{\text{reg}}$.
One can prove that the path integral value does not depend on the coupling constants for such SUSY-exact Lagrangians.
For example, see section 2.3 in \cite{Gang:2009wy} as a lighting discussion.
Then, after decomposing each field as a sum of locus value and quantum fluctuation scaled by the coupling constant as did in \cite{Kapustin:2009kz}, and taking the coupling constant to be $\infty$,
we arrive at a certain Gaussian Lagrangian.
It means all we should know for the quantum computations are so-called one-loop determinants, in other words, Gaussian integrals around the locus \eqref{locus}.
Technically, it is very important to have off-shell SUSY in order to scale out the coupling constants from the resulting path integrals completely.
\paragraph{Yang-Mills part.}
We use $\Ai , \varphi$ as fluctuation modes from locus value $0, \frac{\si_0}{f}$.
After scaling up the Yang-Mills coupling constant, we get Gaussian Lagrangian
\begin{align}
d^3 x \sqrt{g} \mathcal{L}
=&
\frac{1}{2} \Big(
f d \Ai \wedge * d \Ai
+f d^f \varphi \wedge * d^f \varphi
- f[ \frac{\si_0}{f} , \Ai]\wedge[\frac{\si_0}{f} , * \Ai]
+ 2i [\si_0 , \Ai]\wedge * d^f \varphi
\Big) \notag \\
&+
d^3 x \sqrt{g} \ 
 f \ \tr  \Big( i \olam  \g^\mu \D_\mu \lam + \frac{if'}{2f}  \olam \g^3 \lam + i \olam [ \frac{\si_0}{f}, \lam] - \frac{1}{2sf^2}  \olam \lam \Big) 
 .
\end{align}
We define the notation $d^f \varphi = f^{-1} d  (f \varphi)$ for simplicity.
In order to gauge out the cross term between $\Ai$ and $\varphi$,
we take the following gauge condition
\begin{align}
d^{\dagger} (\frac{\Ai}{f}) =0.
\end{align}
In addition, note that the vector kinetic term can be represented by
\begin{align}
f d \Ai \wedge * d \Ai
=
f \Big( \frac{\Ai}{f} \Big) \wedge * [*d f * d f] \Big( \frac{\Ai}{f} \Big),
\end{align}
up to total derivative.
As same, the mass term can be rewritten as
\begin{align}
- f[ \frac{\si_0}{f} , \Ai]\wedge[\frac{\si_0}{f} , * \Ai]
=
- f[ \si_0 , \frac{ \Ai}{f}]\wedge[\si_0 , *  \frac{\Ai}{f}] .
\end{align}
Now, let us define $B= \frac{\Ai}{f}$, then we get the gauge fixed linearized Lagrangian for boson
\begin{align}
d^3 x \sqrt{g} \mathcal{L}_B
=
f \tr \Big( B \wedge *[*df *df + ad^2(\si_0)] B \Big) .
\end{align}
We expand the field in vector multiplet with the Lie algebra basis ($H_i, E_\alpha, E_{-\alpha}$) where $\alpha$ represent positive roots and $H_i$ gives basis for Cartan subalgebra,
and take $\si_0 = \si_0^i H_i$ by using redundancy.
Then the relevant parts depending on $\si_0$ are
\begin{align}
&f  B_{- \alpha} \wedge *[*df *df + \alpha^2(\si_0)] B_\alpha , \label{vecb} \\
&  \olam_{-\alpha} [  i f \g^\mu \D_\mu + \frac{if'}{2} \g^3 + i \alpha(\si_0) - \frac{1}{2sf} ] \lam_\alpha , \label{vecf}
\end{align}
where $\alpha(\si_0)$ is defined by
\begin{align}
[\si_0 , E_\alpha] = \alpha(\si_0) E_\alpha.
\end{align}
The one-loop determinant is expressed by the product
\begin{align}
\prod_{\alpha}\frac{\text{all eigenvalues of \eqref{vecf}}}{\sqrt{\text{all eigenvalues of \eqref{vecb}} |_{\text{ker} d^{\dagger}}}}  \times | \alpha(\si_0) | . \label{v1loop}
\end{align}
Multiplication $\times | \alpha(\si_0) |$ represents contributions from well known hermitian matrix model Vandermonde determinant from fixing $\si_0$.

At first sight, it might look necessary to determine all eigenmodes and their eigenvalues in order to compute \eqref{v1loop}.
However, it is expected that certain modes will be cancelled because of supersymmetry.
In fact, one can check the following eigenmodes pairing structure.
Let $\mathcal{B}$ satisfying $d^{\dagger} \mathcal{B} =0$ is a vector eigenmode, 
\begin{align}
 & M \mathcal{B} = i \alpha(\si_0) \mathcal{B}  - *d f\mathcal{B} \label{veig}.
\end{align}
Then, we can construct a corresponding spinor eigenmode $\Lambda'$ as
\begin{align}
\Lambda' = \g^\mu \e \mathcal{B}_\mu .
\label{sppair}
\end{align}
On the other hand, we can construct a vector eigenmode $\mathcal{B}'$ from
a spinor eigenmode $\Lambda$ which satisfy
\begin{align}
& M \Lambda = (  i f \g^\mu \D_\mu + \frac{if'}{2} \g^3 + i \alpha(\si_0) - \frac{1}{2sf} ) \Lambda , \label{seig}
\end{align}
via
\begin{align}
\mathcal{B}' \equiv f^{-1} \Big( d(f \oep \Lambda) +\big( iM + \alpha(\si_0) \big) \oep \g_\mu \Lambda d x^\mu \Big) . \label{vecpair}
\end{align}
As discussed in \cite{Alday:2013lba, Hama:2011ea, Martelli:2011fu} or in \cite{Drukker:2012sr} via index theorem, the following modes give relevant contribution to the one-loop determinant \eqref{v1loop}.

(1):
The $\mathcal{B}$ which satisfy \eqref{veig} with a condition $\eqref{sppair} =0$.

(2):
The $\Lambda$ which satisfy \eqref{seig} with a condition $\eqref{vecpair}=0$.
\\
We solve the equations for (1) first.
The condition $\eqref{sppair}=0$ can be solved by taking
\begin{align}
\mathcal{B}_1 = \frac{i}{f} \mathcal{B}_3 \sin \theta,
\ \ \ \
\mathcal{B}_2 = \frac{i}{sf} \mathcal{B}_3 \cos \theta.
\end{align}
Then, by defining the 3rd component as
\begin{align}
\mathcal{B}_3 = y(\theta) e^{im\phi - in\chi},
\end{align}
the equation \eqref{veig} gives
\begin{align}
M &= \frac{1}{s}m + n + i \alpha(\si_0), \\
0 &= \Big( \partial_{\theta} -\frac{\sin \theta}{\cos \theta} ( \frac{1}{f} m + 1) + \frac{\cos \theta}{\sin \theta} (\frac{1}{sf} n +1) \Big) y .
\label{eigenv1}
\end{align}
We can restrict the values for $m$ and $n$ through a condition that the $y$ has no singularities at  $\theta \sim 0$ or $\pi/2$.
In fact, one can verify the following behavior for $y$ around $\theta \sim 0$ or $\pi/2$ as
\begin{align}
y \sim (\cos \theta)^{(-m-1)} (\sin \theta)^{(-n-1)}. 
\end{align}
In order to say $y$ is well defined function on $[0, \pi/2]$,
we obtain a constraint $m,n \leq -1$.

Second, we solve the equations for (2) by assuming
\begin{align}
\Lambda
=
\e \Phi_0 + \oep \Phi_2.
\end{align}
It is necessary to introduce R-charge assignments as $\Phi_0$ to 0 and $\Phi_2$ to 2 in order to preserve $\Lambda$'s R-charge.
Here, we take the following form for each $\Phi$
\begin{align}
\Phi_0 = \varphi_0 (\theta) e^{im \phi - i n \chi},
\quad
\Phi_2 = \varphi_2 (\theta) e^{im \phi - i n \chi}.
\end{align}
By substituting them into \eqref{seig} and $\eqref{vecpair}=0$, we get the following equations:
\begin{align}
M= \frac{1}{s} m + n + i \alpha(\si_0), 
\quad
(2sf' + sf \pa_\theta) \varphi_0 &= i(m+ns) \varphi_2, 
\label{sp1}
\\
\Big( 2sf' + sf \pa_\theta + sm \frac{\sin \theta}{\cos \theta} - n \frac{\cos \theta}{\sin \theta} \Big) \varphi_0 &= 0.
\label{sp2}
\end{align}
Well defined behavior for $\varphi_0$ determined by \eqref{sp2} gives $m,n \geq 0$.
In addition, the $m=n=0$ case is excluded.
Once we take so, the 2nd equation for \eqref{sp2} gives $(2sf' + sf \pa_\theta) \varphi_0 = 0 \varphi_2$, then one cannot get the explicit form for $\varphi_2$.

Gathering the each eigenvalues in (1) and (2), the one-loop determinant including the vandermonde determinant becomes
\begin{align}
\eqref{v1loop}
=
\prod_{\alpha} \frac{|\alpha(\si_0)|}{i \alpha(\si_0)} \prod_{m,n \geq 0} \frac{ \frac{1}{s}m+n+i \alpha(\si_0)}{ \frac{(-m-1)}{s} + (-n-1) + i \alpha(\si_0)} 
=
\prod_{\alpha>0} 4 \sinh \big( \pi \alpha( \si_0) \big) \sinh \big( \pi s \alpha (\si_0) \big).
\end{align}
Here $\alpha>0$ means positive roots.
We take zeta function regularization when we convert infinite product into $\sinh$ form.
\paragraph{Matter part.}
We discuss the localization with the matter Lagrangian here.
For later convenience, we use the following $\dl_{\oep}$-exact deformed Lagrangian
\begin{align}
\mathcal{L}_{reg}
&=
\mathcal{L}_{mat}
+
f^{2 \Delta-2} \frac{2(\Delta-1)}{sf^2} \dl_{\oep} \dl_\e \ophi \phi
\notag \\
&=
(2sf^{2\Delta -1}) \Big(
 \D_\mu \ophi \D^\mu \phi 
 -i \frac{2(\Delta -1)}{sf^2} \frac{1}{2sf} \ophi v^\mu \D_\mu \phi 
 +  \ophi \si^2 \phi \color{black}
+ i  \ophi \big( D +\frac{1}{sf^2} \si \big) \phi \color{black}
                          \notag \\ & \qquad \qquad \qquad
+ \frac{2\Delta^2-3\Delta}{2(sf^2)^2}  \ophi \phi   - \frac{\Delta  (2\Delta -1)}{2} (\frac{f'}{f})^2  \ophi \phi  + \frac{\Delta}{4} R  \ophi \phi 
                          \notag \\ & \qquad \qquad \qquad
- i \opsi \g^\mu \D_\mu \psi
-i \frac{f'}{f} (\Delta - \frac{1}{2} ) \opsi \g^3 \psi
 + i \opsi \si \psi 
- \frac{1}{2sf^2} (\opsi \psi) 
                          \notag \\ & \qquad \qquad \qquad
+ \frac{(\Delta -1)}{sf^2}\frac{1}{2sf}(\opsi \g^\mu v_\mu  \psi) 
+i \opsi \lam \phi
- i \ophi \olam \psi
+ \oF F
\Big).
\label{regm}
\end{align}
Integrating by part in \eqref{regm}, and after usual scaling up coupling constant procedure, we get the operators
 $\Delta_\phi$ and $\Delta_\psi$  
\begin{align}
\Delta_\phi
=
&-f^2 \D^2
  - (2\Delta -1) ff' \D_\theta
   - \frac{i(\Delta -1)}{s^2 f} v^\mu \D_\mu
   +\si_0^2
   +  \frac{2\Delta^2-3\Delta}{2(sf)^2}  - \frac{\Delta  (2\Delta -1)}{2} (f')^2
   + f^2 \frac{\Delta}{4} R
, \label{matb} \\
\Delta_\psi
=
& -i f \g^\mu \D_\mu - i f'(\Delta - \frac{1}{2} ) \g^3 + i \si_0 - \frac{1}{2sf} + \frac{(\Delta-1)}{sf} \frac{1}{2sf} \g^\mu v_\mu
. \label{matf}
\end{align}
We can also construct supersymmetric pairing within these fields.
Let $\Psi$ satisfies 
\begin{align}
\Delta_\psi \Psi= M\Psi,
\label{mat1}
\end{align}
then 
\begin{align}
\Phi '=\oep \Psi
\label{matp1}
\end{align}
gives $\Delta_\phi \Phi ' = M(M-2i\si_0)\Phi '$.
We show one easier way to check it in Appendix \ref{B}.

On the other hand, let $\Phi$ satisfies 
\begin{align}
\Delta_\phi \Phi = M(M-2i \si_0)\Phi,
\label{mat2}
\end{align}
then we can construct spinors as
\begin{align}
\Psi_1' = f^{-1} \e \Phi ,
\ \ \ \ 
\Psi_2' = i \g^\mu \e \D_\mu \Phi + i \e \frac{\si_0}{f}  \Phi - \frac{\Delta}{sf^2} \e \Phi - i \Delta \frac{f'}{f} \oep \Phi. \label{Psi12}
\end{align}
These spinors gives the following equation in matrix form
\begin{align}
\begin{pmatrix}
\Delta_\psi \Psi_1 ' \\
\Delta_\psi \Psi_2 '
\end{pmatrix}
=
\begin{pmatrix}
2i \si_0 & -1 \\
-M(M-2i \si_0) & 0
\end{pmatrix}
\begin{pmatrix}
\Psi_1 ' \\
\Psi_2 '
\end{pmatrix}
.
\end{align}
This matrix corresponds to two eigenvalues $M$ and $2i \si_0 - M$ because we concentrate on the determinant only.

Therefore, as same as vector multiplet case, nontrivial contributions to one-loop determinant
\begin{align}
\frac{\text{all eigenvalues of \eqref{matf}}}{\text{all eigenvalues of \eqref{matb}}} \label{m1loop}
\end{align}
come from the following relevant modes:

(3): The $\Psi$ which satisfy \eqref{mat1} with a condition \eqref{matp1} = 0.

(4): The $\Phi$ which satisfy \eqref{mat2} with a condition $\Psi_2 ' = M \Psi_1 '$.
\\
We solve (3) first. Because $\e, \oep$ are linearly independent and $\oep \oep =0$, the condition $\eqref{matp1}=0$ for $\Psi$ gives an expression $\Psi = \oep F$, where R-charge for $F$ must be $2- \Delta$.
The equation \eqref{mat1} is equivalent to the following two equations:
\begin{align}
(M + \frac{\Delta - 2}{sf} - i \si_0 ) F &= - \frac{i}{s} \D_\phi F + i \D_\chi F,
\notag \\
\Big( f \D_\theta +f' (  \Delta +1 ) \Big) F
&=
i\frac{\sin \theta}{\cos \theta} \D_\phi F + \frac{i}{s} \frac{\cos \theta}{\sin \theta} \D_\chi F. \label{mateq}
\end{align}
By assuming $F \propto e^{i(m+[2-\Delta]/2)\phi - i(n+[2-\Delta]/2)\chi}$, we can determine the form of $M$ from the first equation in \eqref{mateq} as
\begin{align}
M= i \si_0 + \frac{m}{s} + n - \frac{\Delta-2}{2} \Big( \frac{1}{s} +1 \Big).
\end{align}
From the second equation in \eqref{mateq}, we can determine the asymptotic behavior for $F$ around $\theta \sim 0 , \pi/2$.
$F$ approaches to
\begin{align}
F \sim \cos^m \theta \sin ^n \theta .
\end{align}
Therefore we must have $m,n \geq 0$ in order to get well defined $F$.

Now, we turn to the calculation for (4).
Let us explain the meaning of the condition $\Psi_2 ' = M \Psi_1 '$, more explicitly,
\begin{align}
i \g^\mu \e \D_\mu \Phi + i \e \frac{\si_0}{f}  \Phi - \frac{\Delta}{sf^2} \e \Phi - i \Delta \frac{f'}{f} \oep \Phi
=
M f^{-1} \e \Phi 
. \label{miss}
\end{align}
This condition guarantees the missing $M$ from spinor eigenvalues because $\Delta_\psi \Psi_2 ' = -M(M-2i \si_0) \Psi_1 '= (2 i \si_0 - M) \Psi_2 '$ and this means that the corresponding eigenvalue is $(2i \si_0 - M)$ alone.
As a result, this scalar eigenvalue $M$ only contribute to one-loop determinant.
We can get relevant equations for computing this unpaired scalar eigenvalue by rewriting \eqref{miss} into the following form,
\begin{align}
(M + \frac{\Delta}{sf} - i \si_0 ) \Phi
&=
- \frac{i}{s} \D_\phi \Phi + i \D_\chi \Phi,
\notag
\\ 
(f \D_\theta +  f'\Delta ) \Phi
&=
- i \frac{\sin \theta} {\cos \theta} \D_\phi \Phi
-  \frac{i}{s} \frac{\cos \theta}{\sin \theta} \D_\chi \Phi
. 
\label{scalareq}
\end{align}
By taking the ansatz $\Phi \propto e^{i(m-\Delta/2) \phi -i(n - \Delta/2) \chi}$,
we can get the form for $M$ from the first equation in \eqref{scalareq} as
\begin{align}
M= i \si_0 - \frac{m}{s} - n - \frac{\Delta}{2} \Big( \frac{1}{s} + 1 \Big)
.
\end{align}
From the second equation in \eqref{scalareq}, we get the constraint on $m,n$ as $
m,n \geq 0$ in order to avoid singular behavior of $\Phi$.

Gathering the results in (3) and (4), we get the one-loop determinant \eqref{m1loop} as
\begin{align}
\prod_{m,n \geq 0} \frac{\frac{m}{s}+n + i \si_0 - \frac{\Delta -2}{2} (\frac{1}{s} +1)}{\frac{m}{s}+n - i \si_0 + \frac{\Delta}{2}(\frac{1}{s}+1)}.
\label{mat1loop1}
\end{align}
We can write it more familiar form by defining
\begin{align}
b \equiv (1/s)^{\frac{1}{2}} ,
\quad
Q \equiv b + b^{-1} ,
\quad
\hat{\si_0} \equiv s^{\frac{1}{2}} \si_0,
\end{align}
as defined in \cite{Hama:2011ea}.
Then, \eqref{mat1loop1} turns to be the double sine function 
\begin{align}
\prod_{m,n \geq 0} \frac{mb+nb^{-1} + \frac{Q}{2} + i \hat{\si_0} + \frac{Q}{2}(1-\Delta)}{mb + n b^{-1} + \frac{Q}{2} - i \hat{\si_0} - \frac{Q}{2}(1-\Delta)}
=s_b \big( \frac{iQ}{2}(1- \Delta) - \hat{\si_0} \big).
\end{align} 

\paragraph{Boundary conditions and background gauge.}
Since the Fourier expansions for $F$ and $\Phi$ contain unusual factors $e^{i[2-\Delta]/2\phi - i[2-\Delta]/2\chi}$ and $ e^{-i\Delta/2\phi + i\Delta/2\chi}$ respectively\footnote{Note the Fourier expansions for $\Lambda$ in vector multiplet calculation are also unusual because our Killing spinors do not have $\phi,\chi$ dependence. Therefore, this argument is true for vector multiplet R-charged fields $\lam, \olam$ too.},
one might be puzzled where such terms come from.
The answer is a gauge condition for background $U(1)$ gauge field $V_\mu$.
Let us regard a field $\varphi$ in our SUSY multiplet.
And define $\tilde{\varphi}$ as the corresponding field in \cite{Hama:2011ea}.
Then the transformation rule from $\varphi$ to $\tilde{\varphi}$ is as follows
\begin{align}
\varphi = e^{i \frac{\hat{q}}{2} (\phi - \chi)} \tilde{\varphi}
,
\end{align}
where $\hat{q}$ is R-charge.
This relation is valid with Killing spinors too.
This procedure changes the background gauge field in Lagrangians as
\begin{align}
V \to \tilde{V}= V - \frac{1}{2} d ( \phi - \chi).
\end{align}
Second term comes from $\pa_\mu e^{i \frac{\hat{q}}{2} (\phi - \chi)}$ in the kinetic terms.
Then, the boundary conditions for fields $\tilde{\varphi}$ and the background field $\tilde{V}$ become as same ones in \cite{Hama:2011ea}.
\paragraph{Chern-Simons term.}  \label{sec;wil}
Here, we would like to consider the value of classical Chern-Simons action on the locus \eqref{locus}.
We get same value with squashed sphere as follows
\begin{align}
S_{CS}
&=
\frac{ik}{4 \pi} \int d^3 x \ \mathcal{L} _{CS} (A=\lam= \olam= 0, \si = \frac{\si_0}{f}, D= - \frac{\si_0}{sf^3} ) \notag \\
&= 
\frac{ik}{4 \pi} \tr \si_0^2 \int_{0} ^{\pi/2} \sin \theta \cos \theta d \theta \int_0^{2 \pi} d \phi \int_0^{2 \pi} d \chi \  ( -2 \frac{1 }{sf^4} ) \notag  \\
&= 
-ik \pi s \tr \si_0^2 .
\end{align}
\paragraph{Wilson loop.}
Let us define the supersymmetric Wilson loop as 
\begin{align}
W (R,C;A, \si)
=
\tr _R \mathcal{P}  \exp \Big( \oint_C d \tau (i A_\mu \dot{x}^\mu + \si |\dot{x}| \Big) ,
\end{align}
where $R$ is an arbitrary representation of the gauge group, $\tau$ parametrizes the integral contour $C$ via $x^\mu = x^\mu (\tau)$ and $|\dot{x}| = \sqrt{\dot{x}_\mu \dot{x}^\mu}$.
We can use localization method in order to calculate Wilson loop if and only if the following condition satisfied
\begin{align}
\dl_{\oep} W (R,C;A, \si) = 0 .
\end{align}
This constraint reduces to
\begin{align}
\oep(\g_\mu \dot{x}^\mu +|\dot{x}|) = 0.
\end{align}
One can find easily this is equivalent to the following ODE
\begin{align}
\dot{x}^\mu \frac{\pa}{\pa x^\mu}
=
\left\{ \begin{array}{ll}
\frac{|\dot{x}|}{f(\theta)} \Big(\frac{1}{s} \frac{\pa}{\pa \phi} -  \frac{\pa}{\pa \chi} \Big) & \quad (\theta \neq  0, \frac{\pi}{2} ) \\
|\dot{x}| \frac{\pa}{\pa \phi}  & \quad (\theta =  0 ) \\
- |\dot{x}| \frac{\pa}{\pa \chi} & \quad (\theta =\frac{\pi}{2} ) \\
\end{array} \right. . \label{1/2BPSWl}
\end{align}
It is useful to introduce auxiliary variables $l, \tilde{l}$ as
\begin{align}
s= \frac{l}{\tilde{l}} \
.
\end{align}
If and only if the value $s$ is a rational number, we can get periodic contours which satisfy the differential equation in \eqref{1/2BPSWl} for $\theta \neq 0, \frac{\pi}{2}$.
We assume $l , \tilde{l}$ are co-prime integers.
In each case, classical value for Wilson loop insertions are
\begin{align}
W(R, C_\theta ; 0, \frac{\si_0}{f(\theta)})
=
\left\{ 
\begin{array}{ll}
\tr_R e^{2 \pi l \si_0} & \quad (\theta \neq  0, \frac{\pi}{2} )  \\
\tr_R e^{2 \pi \frac{l}{\tilde{l}} \si_0} & \quad (\theta = 0 )  \\
\tr_R e^{2 \pi \si_0} & \quad (\theta = \frac{\pi}{2} ) 
\end{array} \right. .
\end{align}
The topological shape turns to be so-called $(l,\tilde{l})$-torus knot when $\theta \neq 0, \frac{\pi}{2}$ through as same discussion in \cite{Tanaka:2012nr}.
And the expectation values become knot invariants if we consider vector multiplet only.

\section{Concluding Remarks}\label{sec;remark}
In this paper we presented supersymmetric gauge theories on round sphere with unfamiliar Killing spinors, and performed supersymmetric localization.
Calculations in this paper is inspired by the results on gravity side \cite{Martelli:2011fu}.
They asked whether the localization and field theory partition function are invariant under Weyl rescalings
and conjectured that this will be true at least for large N because such geometrical description comes from different slicings of $AdS_4$.
Our results answer to the question in \cite{Martelli:2011fu} with finite N.
As discussed by them, the corresponding gravity solution can be lift to the eleven dimensional supergravity solution $AdS_4 \times_{\text{twisted}}$ Sasaki-Einstein 7-manifold ($SE_7$).
It is interesting to compare the knotted Wilson loop's expectation value at large $N$ with 
the classical M2-brane action.
For example, \cite{Farquet:2013cwa} calculate M2-brane action by using worldvolume SUSY preserving condition and
 argue the size (and position in $SE_7$) of M-theory circle wrapped by this M2-brane within $AdS_4 \times SE_7$ setup.
 In our case, it is expected that the knotted closed string is emitted to the bulk of $AdS_4$.
It may be interesting to study how the M-theory circle wrapped by M2-brane is lifted to 11-dimensional space when we lift this ``knotted surface" onto the $AdS_4 \times_{\text{twisted}} SE_7$ by following their argument.
Our results ensure the use of the usual $AdS_4$ patch when one calculate the minimal surface.

One puzzle may be the meaning of another supersymmetry on round $S^3$.
It is basically Weyl transformed one as we noted, however it means we can take different supersymmetries on the same Riemannian manifold.
As studied by many authors \cite{Closset:2012ru, Alday:2013lba, Nian:2013qwa}, this fact may suggest that
supersymmetry on curved spaces do not depend only on the metric, but depend on other structures.
It is interesting to study the cases in other dimension.

Another interesting application can be found in \cite{Honda:2012ni, Asano:2012gt}.
Our construction, one-parameter deformed SUSY on round sphere, guarantees the use of invariant one-forms when we write down the Lagrangians.
This may suggest the possibility for constructing a proof of Large N reduction for one-parameter deformed theories.

\acknowledgments

I would like to thank S. Yamaguchi, K. Hashimoto, H. Mori for valuable discussions and comments. And I would also like to thank S. Terashima, K. Yoshida, S. Sugishita, K. Ohta, K. Hosomichi, N. Hama, M. Honda, Y. Asano, and J. Schmude for various comments and K.Oda and N. Iizuka for encouragement.
The work of A.T. was supported in part by JSPS fellowships for Young Scientists.

\paragraph{Note added.} 
We showed that it is possible to get nontrivial results even on round sphere.
This fact is reported in \cite{Nian:2013qwa} too.

\paragraph{Note added.} 
We comment on other dimension's case in the concluding section.
Then, \cite{Closset:2013vra} which studies 4D cases and 3D cases appears.


\appendix
\section{Formulas} \label{A}
\paragraph{Killing spinor bilinears.}
Following equations are valid
\begin{align}
\e \e =0 , 
\quad
\oep \oep =0, &
\quad 
\oep \e = - \e \oep = 2sf,
\\
\e \g^a \e = (2is \sin \theta, 2i \cos \theta, 2sf),
\quad &
\oep \g^a \oep = (2is \sin \theta, 2i \cos \theta, -2sf),
\\
\oep \g^a \e = \e \g^a \oep = (- & 2  \cos  \theta, 2s \sin \theta, 0).
\end{align}

\paragraph{Key formulas.}
First ones describe some relations between $\g V$ type contracted gamma matrices and Killing spinors.
\begin{align}
&-i \frac{f'}{f} \g^\mu V_\mu \e =  \frac{i}{2} \g^{\mu \nu} V_{\mu \nu} \oep = + \frac{1}{2}  (\frac{f'}{f})' \oep + i \frac{f'}{sf^3} \e , \\
&-i \frac{f'}{f} \g^\mu V_\mu \oep =  \frac{i}{2} \g^{\mu \nu} V_{\mu \nu} \e = - \frac{1}{2}  (\frac{f'}{f})' \e - i \frac{f'}{sf^3} \oep .
\end{align}
Second ones are relations between $(\pa v) \g$ type matrices and Killing spinors.
\begin{align}
& \frac{i}{2} \frac{1}{s^2 f^3} (\pa_\mu v_\nu) \g^{\mu \nu} \oep  =+ 2 (\frac{1}{sf^2})^2\oep + 2i \frac{f'}{sf^3} \e, \\
& \frac{i}{2} \frac{1}{s^2 f^3} (\pa_\mu v_\nu) \g^{\mu \nu} \e  = - 2 (\frac{1}{sf^2})^2\e - 2i \frac{f'}{sf^3} \oep.
\end{align}
Third ones are $v \g$ types as follows
\begin{align}
\frac{1}{2sf} v_\mu \oep \g^\mu = \oep,
\quad
\frac{1}{2sf} \g^\mu \e v_\mu = \e.
\end{align}
In addition, one can verify
\begin{align}
1=
\Big( \frac{1}{sf^2} \Big)^2
-
\Big( \frac{f'}{f} \Big)^2
-
\Big( \frac{f'}{f} \Big)'
\quad
.
\end{align}

These formulas are used to compute consistency of the SUSY algebra, exactness of the Lagrangians, and one-loop determinants.
\begin{table}[tbp]
\begin{tabular}{|c||c|c|} 
\hline
Killing spinor & $\e$ & $\overline{\e}$\\ \hline \hline
spin & 1/2 & 1/2  \\ \hline
$\hat{q}$ & +1 & $-1$  \\ \hline
\end{tabular}
\centering
\begin{tabular}{|c||c|c|c|c|c||c|c|c|c|c|c|} 
\hline
Field & $A_{\mu}$ & $\si$ & $\lam$ & $\olam$ & $D$  & $\phi$ & $\overline{\phi}$ & $\psi$ & $\overline{\psi}$ & $F$ & $\overline{F}$ \\ \hline \hline
spin & 1 & 0 & 1/2 & 1/2 & 0  & 0 & 0 & 1/2 & 1/2 & 0 & 0 \\ \hline
$\hat{q}$ & 0 & 0 & +1 & $-1$ & 0 & $-\Delta$ & $\Delta$ & $-(\Delta - 1)$ & $\Delta-1$ & $-(\Delta-2)$ & $\Delta-2$ \\ \hline
\end{tabular}
\caption{\label{tab:A1}}
\end{table}

\paragraph{Covariant derivative.}
Before using linearized Lagrangians, we use 
\begin{align}
\D_\mu = \nabla _\mu -i \rho(A_\mu) - i \hat{q} V_\mu, \label{eq:cova1}
\end{align}
where  $\nabla_\mu$ is usual derivative or spin covariant derivative or vector covariant derivative respectively corresponding to the field spin,
$\rho$ is adjoint representation or general representation for fields in vector or matter multiplet and zero for Killing spinors.
$V_\mu$ is defined in \eqref{bg} and $\hat{q}$ is the background $U(1)$ charge defined as in Table~\ref{tab:A1}.
\section{Proof for paired eigenmodes}\label{B}
Here, we check 
\begin{align}
\Delta_\psi \Psi = M \Psi
\quad
\to
\quad
\Delta_\phi (\oep \Psi)
=
M(M-2i \si_0) (\oep \Psi)
\end{align}
by the use of Weyl transformation from squashed sphere.
Let us define \textit{hyperbolic squashed sphere} metric 
\begin{align}
d \tilde{s}^2 
= 
\frac{d \ttheta ^2}{h^2(\ttheta)}
+\cos^2 \ttheta d \tilde{\phi}^2
+ \frac{1}{s^2} \sin^2 \ttheta d \tilde{\chi}^2
,
\label{sqmet}
\end{align}
where $h^2(\ttheta)=s^2 \cos^2 \ttheta + \sin^2 \ttheta$.
One can verify the relation $f(\theta) = \frac{1}{h(\ttheta)}$ easily.
Here, we quote some results in \cite{Martelli:2011fu}.
\paragraph{Covariant derivative.}
We define a covariant derivative as
\begin{align}
\tilde{\D}_\mu = \tilde{\nabla}_\mu - i \hat{q} V_\mu
\end{align}
where $\tilde{\nabla}$ is constructed from \eqref{sqmet}.
\paragraph{Operators on squashed sphere.}
\begin{align}
\tilde{\Delta}_\phi
&=
- \tilde{\D}_\mu \tilde{\D}^\mu
- 2i (\Delta -1) h(\ttheta) \frac{\tilde{v}^\mu}{2}  \tilde{\D}_\mu
+ \tilde{\si}^2_0
+\frac{2\Delta^2 - 3 \Delta}{2} h^2(\ttheta)
+ \frac{\Delta}{4} \tilde{R},
\label{sqbos}
\\
\tilde{\Delta}_\psi
&=
-i \tilde{\g}^\mu \tilde{\D}_\mu 
+ i \tilde{\si}_0
- \frac{h(\ttheta)}{2}
+(\Delta-1) h(\ttheta) \tilde{\g}^\mu \frac{\tilde{v}_\mu}{2} ,
\end{align}
where $\tilde{v}_\mu = \frac{1}{sf(\theta)} v_\mu$.
Now, one can show the following statement
\begin{align}
\tilde{\Delta}_\psi \tilde{\Psi}
= \tilde{M} \tilde{\Psi}
\quad
\to
\quad
\tilde{\Delta}_\phi (\tilde{\oep} \tilde{\Psi})
=
\tilde{M} (\tilde{M} - 2i \tilde{\si}_0) (\tilde{\oep} \tilde{\Psi}),
\label{onsq}
\end{align}
where $\tilde{\oep} = (sf)^{-1/2} \oep$ as one can find in \cite{Martelli:2011fu}.
\paragraph{Proof.}
According to \cite{Hama:2010av}, the field $\Psi$ scales as
\begin{align}
\tilde{\Psi} = (sf)^{\Delta + 1/2} \Psi,
\end{align}
Then, 
\begin{align}
(\tilde{\oep} \tilde{\Psi}) = (sf)^\Delta (\oep \Psi)
.
\end{align}
Each terms in \eqref{sqbos} can be mapped into the corresponding operators on \textit{round sphere}
as follows.
\begin{align}
&- \tilde{\D}_\mu \tilde{\D}^\mu (sf)^\Delta (\oep \Psi)
\notag \\
&=
(sf)^\Delta
\Big(
- (sf)^2 \D_\mu \D^\mu
- (sf)^2 (2\Delta-1) \frac{f'}{f} \D_\theta 
+ (sf^2) \Delta \big( - (\frac{f'}{f})' + (\frac{f'}{f})^2 \big)
- (sf^2) \Delta^2 (\frac{f'}{f})^2
\Big)
(\oep \Psi).
\label{1}
\end{align}
\begin{align}
&
\Big(
- 2i (\Delta -1) h(\ttheta) \frac{\tilde{v}^\mu}{2} \tilde{\D}_\mu
+ \tilde{\si}^2_0
+\frac{2\Delta^2 - 3 \Delta}{2} h^2(\ttheta)
\Big)
(sf)^\Delta (\oep \Psi)
\notag \\
&=
(sf)^\Delta  
\Big(
- 2i (\Delta -1) \frac{1}{f(\theta)} \frac{v^\mu}{2} \D_\mu
+ \tilde{\si}^2_0
+\frac{2\Delta^2 - 3 \Delta}{2} \frac{1}{f^2(\theta)}
\Big)
 (\oep \Psi).
 \label{2}
\end{align}
\begin{align}
&\frac{\Delta}{4} \tilde{R}(sf)^\Delta (\oep \Psi)
\notag \\
&=
(sf)^\Delta\frac{\Delta}{4} \tilde{R} (\oep \Psi)
\notag \\
&=
(sf)^\Delta\frac{\Delta}{4} 
(sf)^2
\Big(
R 
+ 4 (\frac{f'}{f})'
-2 (\frac{f'}{f})^2
\Big)
(\oep \Psi).
\label{3}
\end{align}
We used formulas for Weyl scaled laplacian and scalar curvature.
Gathering \eqref{1}, \eqref{2}, \eqref{3}, we get
\begin{align}
&\tilde{\Delta}_\phi (\tilde{\oep} \tilde{\Psi})
\notag \\
&=
(sf)^\Delta s^2
\Delta_\phi \Big| _{\si_0 = \tilde{\si}_0/s}
(\oep \Psi).
\label{weylbos}
\end{align}
Then, by combining the results \eqref{onsq} and \eqref{weylbos},
we arrive at the end of the proof as follows.
\begin{align}
s^2
\Delta_\phi \Big| _{\si_0 = \tilde{\si}_0/s}
(\oep \Psi)
&=
\tilde{M} (\tilde{M} - 2i \tilde{\si}_0) (\oep \Psi)
\notag \\
\leftrightarrow
\notag \\
\Delta_\phi \Big| _{\si_0 = \tilde{\si}_0/s}
(\oep \Psi)
&=
\frac{\tilde{M}}{s} ( \frac{\tilde{M}}{s} - 2i \si_0) (\oep \Psi)
\notag \\
\leftrightarrow
\notag \\
\Delta_\phi \Big| _{\si_0 = \tilde{\si}_0/s}
(\oep \Psi)
&=
M ( M - 2i \si_0) (\oep \Psi),
\end{align}
where we define $\frac{\tilde{M}}{s} = M$.


\providecommand{\href}[2]{#2}\begingroup\raggedright\endgroup

\end{document}